\documentclass[a4paper,twocolumn,11pt,accepted=2024-02-14]{quantumarticle}
\pdfoutput=1

\pdfoutput=1
\usepackage[utf8]{inputenc}
\usepackage[english]{babel}
\usepackage[T1]{fontenc}
\usepackage{amsmath}
\usepackage{hyperref}

\usepackage{hyperref}
\usepackage{multirow}
\usepackage{mathrsfs,amsmath} 
\usepackage{verbatim}
\usepackage{xcolor}
\usepackage{lipsum}
\usepackage{dsfont}
\usepackage{footnote}
\usepackage{amsfonts}
\usepackage{tcolorbox}
\usepackage{mathtools}
\usepackage{amsthm}

\definecolor{shadecolor}{rgb}{0.8,0.9,1}
\newcommand{\ket}[1]{| {#1} \rangle} % for Dirac bras
\newcommand{\bra}[1]{\langle {#1} |} % for Dirac kets
\newcommand{\braket}[2]{\langle {#1} \vphantom{#2} | {#2} \vphantom{#1} \rangle} % for Dirac brackets
\DeclareDocumentCommand{\Tr}{m m O{\big}}{{\rm Tr}_{\:\!{#1}}#3({#2}#3)}

\newtheorem{theorem}{Theorem}

\newtheorem{lemma}{Lemma}
\newtheorem{conj}{Conjecture}

\begin{document}
\title{Towards a measurement theory in QFT: ``Impossible" quantum measurements are possible but not ideal}
%\title{What nonlocal quantum measurements are nonsignaling?}

\author{Nicolas Gisin}
\affiliation{Group of Applied Physics, University of Geneva, 1211 Geneva, Switzerland}
\affiliation{Constructor University, Geneva, Switzerland}

\author{Flavio Del Santo}
\affiliation{Group of Applied Physics, University of Geneva, 1211 Geneva, Switzerland}
\affiliation{Constructor University, Geneva, Switzerland}

%\date{{21.~February  2041}}

\begin{abstract}
\noindent 
Naive attempts to put together relativity and quantum measurements 
%(in combination with the lack of a theory of measurement in QFT) l
lead to signaling between space-like separated regions. In QFT, these are known as \textit{impossible measurements}. We show that the same problem arises in non-relativistic quantum physics, where joint nonlocal measurements (i.e., between systems kept spatially separated) in general lead to signaling, while one would expect no-signaling (based for instance on the \textit{principle of no-nonphysical communication}). This raises the question: Which nonlocal quantum measurements are physically possible? We review and develop further a non-relativistic quantum information approach developed independently of the impossible measurements in QFT, and show that these two have been addressing virtually the same problem. The non-relativistic solution shows that all nonlocal measurements are \textit{localizable} (i.e., they can be carried out at a distance without violating no-signaling) but they (i)  may require arbitrarily large entangled resources and (ii)  cannot in general be \textit{ideal}, i.e., are not immediately reproducible. These considerations could help guide the development of a complete theory of measurement in QFT.

\end{abstract}

\maketitle
%%%%%%%%%%%%%%%%%%%%%%%%%%

%%%%%%%%%%%%%%%%%%%%%%%%%%%%%%%%%%%%%%%%%

\section{Introduction}
\label{intro}

Like every empirical science, physics owns its success to the possibility of inquiring nature, i.e., performing measurements. Although we are about to celebrate the centenary of quantum physics, the most empirically successful theoretical framework ever, there are still fundamental problems when it comes to understanding measurements. On the one hand, there is no consensus on the solution of the \textit{measurement problem}, i.e., what interactions constitute a measurement and what physical mechanism (if any at all) determines the appearance of a single outcome. (Note that producing an outcome in the form of a piece of classical information is \textit{the} main feature of a measurement). On the other hand, and this will be the main focus of this paper, there is little discussion on whether any theoretically acceptable quantum measurement (including of nonlocal variables) can be physically implemented, namely, such that the implementation does not imply the violation of any known physical principles, like, e.g., the no-signaling principle. 

The issue of which quantum measurements are physically possible was firstly raised in the attempt of putting together quantum physics with relativity. Intuitively the problems stem from the following tension. On the one hand, the most fundamental principle of relativity states that signals cannot travel faster than a maximum speed (incidentally that of light). Quantum measurements, on the other hand, involve the  ``collapse" of the state vector at the time of a measurement, which (when considering entangled states of multipartite distant systems) seems to impose an instantaneous change of the state, possibly in a space-like separated region. It so became manifest quite early that relativity and quantum measurements do not easily coexist \cite{landau1931erweiterung, schilpp1949library, hellwig1970formal, aharonov1980states, aharonov1981can, guerreiro2012single}. 

The theory that extends quantum mechanics to a relativistic framework is quantum field theory (QFT). Therein, all the problems with distant systems seem solved by the assumption of \textit{microcausality}, i.e., the algebras of operators defined on any two space-like separated regions commute. However, QFT still lacks to date a complete theory of measurement (i.e., one that yields measurement outcomes and it is therefore able to explicitly model all known quantum phenomena), an issue that has been called “a major scandal in the foundations of quantum physics” \cite{earman}. One may then try to naively apply the  standard rules of quantum measurements in the context of QFT. However, it was pointed out by Sorkin  already in 1993 \cite{sorkin1993impossible} that this leads to a violation of no-signaling between space-like separated regions (provided that measurements take place on finite regions of space-time). These measurements that violate no-signaling are commonly referred to as \textit{impossible measurements} (see section \ref{imp}). 
%%%%%%%%%%%%%%%%%%%%%%%%%%%%%%%%
\begin{figure}[ht]
    \centering
    \includegraphics[width=.48\textwidth]{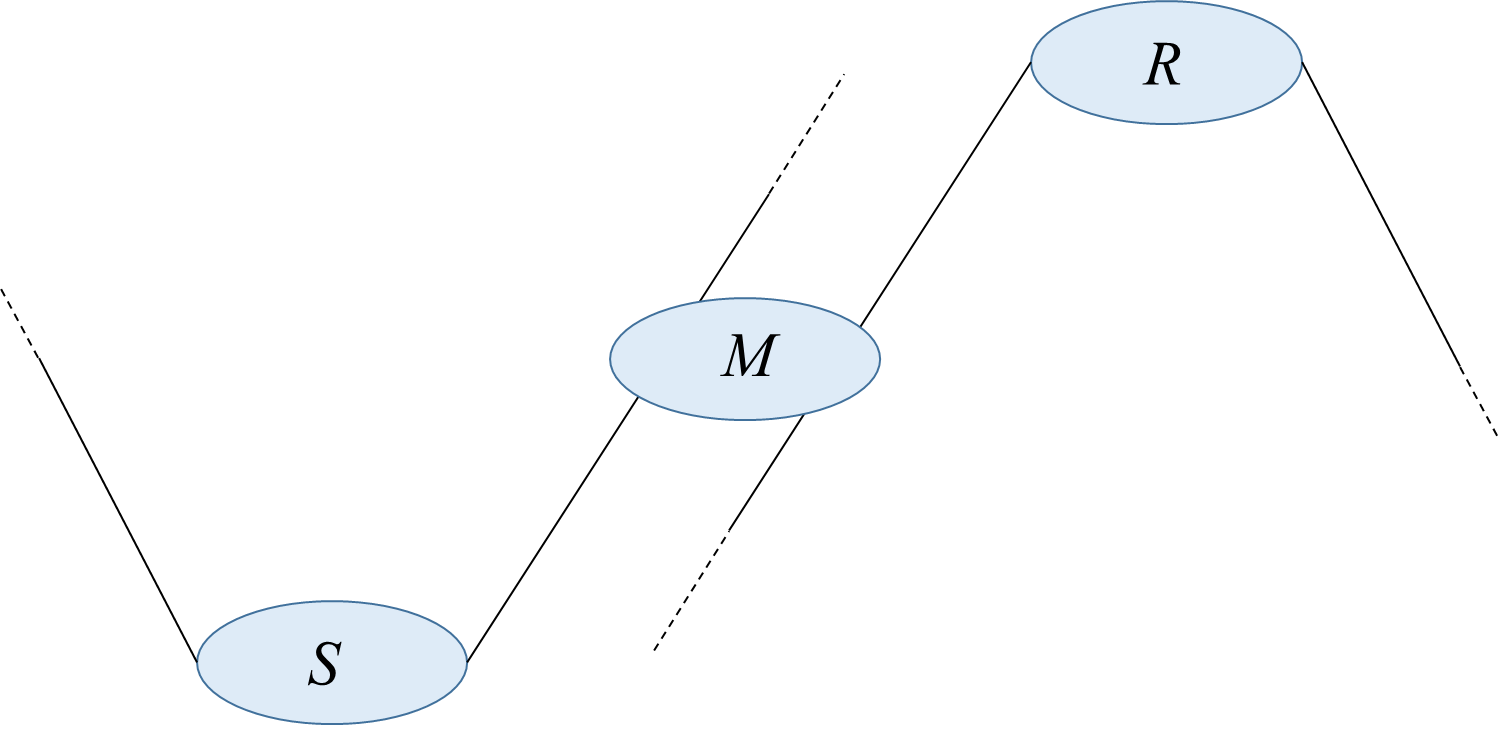}
    \caption{Spacetime diagram of Sorkin's impossible measurement scenario in QFT.}
    \label{fig1}
\end{figure}
%%%%%%%%%%%%%%%%%%%%%%%%%%%%%%%%

The problem of impossible measurements in QFT has gained interest in recent years (for a historical review, see \cite{fraser2023note, papageorgiou2023eliminating}) and two main approaches have been proposed to solve this conundrum: The first one, pursued in Refs. \cite{borsten2021impossible, jubb2022causal, albertini2023ideal}, accepts that microcausality is not enough to single out the physical measurements and puts forward further constraints. The second approach, advanced in Refs. \cite{fewster2020quantum, fewster2020generally, bostelmann2021impossible, fewster2023measurement}, finally provides  QFT with an explicit measurement scheme (by coupling the quantum fields to a probe, which is in turn another quantum field) such that Sorkin's impossible measurements do not seem possible to arise in the first place. 

While these proposed solutions represent a valuable advancement in  understanding  relativistic quantum measurements, they are completely rooted in the formalism of QFT. Yet, it should be noted that impossible measurements are not a characteristic problem of QFT, but rather have a more general scope. In fact, also in non-relativistic quantum theory one may ask what nonlocal joint measurements (i.e., performed on multipartite systems located at a distance from each other) lead to signaling. Studying this problem in a non-relativistic quantum framework is at the same time  simpler and  more general. It is simpler because standard quantum mechanics is provided with a well-established theory of measurement, and one can thus look for the limitations that no-signaling imposes on the theoretically allowed measurements. It is more general because no-signaling could be studied at a more abstract level, in principle without resorting to any physical theory -- in particular to relativity -- to motivate it. In fact, it is true that the principles of  relativity theory represent the most striking justification to impose no-signaling between distant regions, but this is in the end only \textit{one possible} physical justification of why one would impose no-signaling. One can think of different physical motivations, e.g., a thick wall (or any other impenetrable potential barrier), or an interruption in the communication channel. These scenarios exemplify the general \textit{principle of no-nonphysical communication}, which states that communication is possible only when information is encoded into a physical carrier (a letter, a particle, a radio wave, etc.) which physically transports the information from the sender to the receiver \cite{gisin2014quantum}.

Interestingly, the problem of what nonlocal quantum measurements lead to signaling (using a non-relativistic quantum information formalism) has    already been studied in a series of papers \cite{aharonov1986measurement, popescu1994causality, groisman2001nonlocal, groisman2002measurements, vaidman2003instantaneous, groisman2003instantaneous, clark2010entanglement,  beigi2011simplified, gonzales2019bounds}. Therein  it is shown that the ideal measurements (for definitions on different types of measurements, see section \ref{defmeas}) of nonlocal variables (which do not lead to signaling) are those that completely erase the information from the local states (i.e., the post-measurement updated state locally reduces to the identity matrices). We borrow the useful terminology from Ref. \cite{beckman2001causal}, although with a different specific meaning, and we call these measurements \textit{localizable}.\footnote{In Ref. \cite{beckman2001causal}, they originally called those measurements causal \textit{and} localizable, but we emphasize that if a measurement is localizable in the sense used here, this implies that it does not signal.} As we shall see (in section \ref{localiz}), these measurements are almost never ideal, i.e., the updated post-measurement state is in general not an eigenstate of the measurement operator and cannot therefore be immediately reproduced. 
%For example, for two qubits, the only localizable and ideal nonlocal measurement is the Bell state measurement (BSM), besides trivial product measurements.

Note that what Refs. \cite{beckman2001causal} and \cite{fewster2020quantum} call a measurement actually yields no outcomes. In the former, in particular, they merely diagonalize the density matrix in the measurement eigenbasis, which is equivalent to carry out an ideal measurement and ignoring the outcomes (this defines what is usually called a quantum measurement channel). However, if a measurement is not ideal, it does not in general define a quantum channel. Throughout this paper, we will assume that a measurement is \textit{defined} by yielding a single outcome with the probability given by the Born rule (see section \ref{defmeas} for specific definitions).

In doing so, we will build on the series of results in Refs. \cite{aharonov1986measurement, popescu1994causality, groisman2001nonlocal, groisman2002measurements, vaidman2003instantaneous}. These works,  however, received relatively little attention and were developed independently of Sorkin's impossible measurements. Moreover, these two theoretical frameworks -- despite addressing incredibly similar, if not exactly the same, problems -- have never been connected so far.
It is the aim of this paper to review the main results of this non-relativistic quantum information approach, developing further some of the proposed protocols, and  explicitly show how this satisfactorily addresses the problem of Sorkin's impossible measurements in the non-relativistic limit. This will allow us to better understand the theory of nonlocal quantum measurements in general, and hopefully to help guide the development of a complete theory of measurement in QFT.

%Here we review two approaches to the latter problem, aimed at showing that ideal quantum measurements instantaneously carried out over joint systems or extended regions of space-time lead in general to violation of causality and should therefore be rejected on the grounds of physical principles. 

\section{Impossible measurements in QFT and in the non-relativistic limit}
\label{imp}

Let us start by restating Sorkin's impossible measurement scenarios in QFT \cite{sorkin1993impossible}. Let $S$ (for Sender), $M$ (for Middle), and $R$ (for Receiver) be three bounded regions of (Minkowski)  spacetime, arranged as in Fig.\ref{fig1}, each provided with a local algebra     $\mathfrak{U}(S)$,     $\mathfrak{U}(M)$, and     $\mathfrak{U}(R)$, respectively. Since $S$ and $R$ are space-like separated, microcausality imposes that all the operators of one region commute with the ones of the other, i.e., $[\hat S_i,\hat R_j]=0$, $\forall \hat S_i\in \mathfrak{U}(S)$ and $\forall \hat R_j \in \mathfrak{U}(R)$. This supposedly ensures that no operation performed in region $S$ can signal to region $R$.

 \begin{figure}[ht]
    \centering
    \includegraphics[width=.48\textwidth]{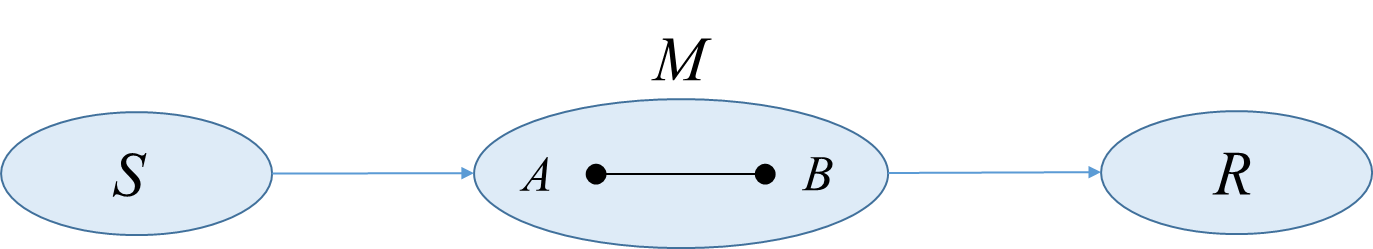}
    \caption{Non-relativistic limit of a Sorkin's impossible measurement scenario.}
    \label{fig2}
\end{figure}

Let us now consider, however, the third  region $M$, which is chosen to be partly in the causal future of $S$ and partly in the causal past of $R$.\footnote{Note that $M$ needs to be an extende region, i.e., to have a finite size. For any such region, however, it is always possible to find two regions $S$ and $R$ arranged in the way just described.} 
Consider now the following protocol. The Sender, situated in region $S$, applies to the state of the field, $\ket{\psi_0}$, a ``kick" operator, i.e., a unitary element of the local algebra $\hat S(\gamma) \in \mathfrak{U}(S)$, which depends on the parameter $\gamma$. 
%This transforms the state of the field as $\ket{\psi_0} \rightarrow A(\lambda)\psi_0:= \ket{\psi_1}$. 
The Middle observer in region $M$ performs a nonselective measurement -- i.e., one which also produces outcomes but they are ignored (i.e., traced over) -- of the observable $\hat \Omega^{(M)}=\{\hat \Pi^{(M)}, \mathds{1} - \hat \Pi^{(M)}\}  \in \mathfrak{U}(M)$, where $\hat \Pi^{(M)}$ is a projector.
Finally, the Receiver, in region $R$, measures the expectation value of an observable $\hat R \in \mathfrak{U}(R)$. The application of the state-update rule after the measurement in $M$, yields the result:
\begin{eqnarray}
       \langle \hat R \rangle &=& \bra{\psi_0} \hat S^*(\gamma) \hat \Pi^{(M)} \hat R \hat \Pi^{(M)} \hat S(\gamma) \ket{\psi_0} \\
       &&+  \bra{\psi_0} \hat S^*(\lambda) (\mathds{1} - \hat \Pi^{(M)}) \hat R (\mathds{1} - \hat \Pi^{(M)}) \hat S(\gamma) \ket{\psi_0}. \nonumber
\end{eqnarray}
  It is easy to show that, given the state-updated after the projective measurement in $M$, in general $\langle \hat R \rangle$ depends on the ``kick" parameter $\gamma$ and therefore the Sender can signal to the Receiver thanks to the  intervention of the Middle observer's measurement (see a simple example in section \ref{example}).

Let us turn to the non-relativistic scenario, as depicted in Fig. \ref{fig2}. In this case, $S$, $M$, and $R$ are three separate laboratories, each provided with a system associated to a suitable Hilbert space, $\mathcal{H}_S$,  $\mathcal{H}_M$, and  $\mathcal{H}_R$, respectively. The systems in $S$ and $R$ are completely arbitrary and each of them can be, in the simplest case,  a qubit (i.e., $\mathcal{H}_S=\mathbb{C}^2=\mathcal{H}_R$). However, we model the fact that $M$ needs to be an extended region of space by it being composed of two spatially separated qubits, $A$ and $B$, such that $\mathcal{H}_M=\mathbb{C}^2 \otimes \mathbb{C}^2$. Following step by step the construction of Sorkin's scenario in QFT, we allow, in this order, $S$ to communicate with $A$, and $B$ to communicate with $R$, whereas no information can be exchanged between $A$ and $B$. This resembles the relativistic case in which $M$ partly overlaps with the future causal cone of $S$ and with the past causal cone of $R$, allowing therefore the (ordered) communication.

We now want to investigate what operations can be performed without violating the assumption of no-signaling between $S$ and $R$ (in QFT this is due to space-like separation, but as explained in section \ref{intro}, one can motivate this without resorting to spacetime with the principle of no-nonphysical communication).
Exactly in the same fashion as the relativistic scenario, assume that at time $t_0$ the global state is $\ket{\psi_0}$ and the Sender performs the local  unitary ``kick" $U^{(S)}$ on her system in $S$. At time $t_1 >t_0$, the Middle observer performs a joint nonselective measurement $\Omega^{(M)}$ on the qubits $A$ and $B$. Finally, at $t_2 > t_1$, the Receiver measures a local observable $\Lambda^{(R)}$, yielding the outcome $\lambda$. The no-signaling constraint is respected if the application of $U^{(S)}$ cannot influence the measurement statistics at $R$, i.e.,  
\begin{eqnarray}
\label{nsnorel}
\text{Prob}(\Lambda^{(R)}=\lambda \text{ at } t_2 | \ket{\psi_0} \text{ at } t_0; \Omega^{(M)} \text{at } t_1)=\\
\text{Prob}(\Lambda^{(R)}=\lambda \text{ at } t_2 | U^{(S)}\ket{\psi_0} \text{ at } t_0; \Omega^{(M)} \text{at } t_1).\nonumber
\end{eqnarray}
Like for Sorkin's impossible measurements in QFT, it is straightforward to show  that the no-signaling condition in Eq. \eqref{nsnorel} is violated by a standard measurement (see section \ref{example}).

As we have already noted, both in the QFT and in the non-relativistic descriptions, the communication between $S$ and $M$ and between $M$ and $R$ are actually allowed, therefore, nothing problematic can stem from there. So the source of the problem fully lies in the (nonlocal) measurement performed at $M$.
Hence, before going into the main argument, let us  sharpen some useful definitions about measurements. 

\subsection{Definitions about measurements}
\label{defmeas}
\begin{figure}[ht]
    \centering
    \includegraphics[width=.44\textwidth]{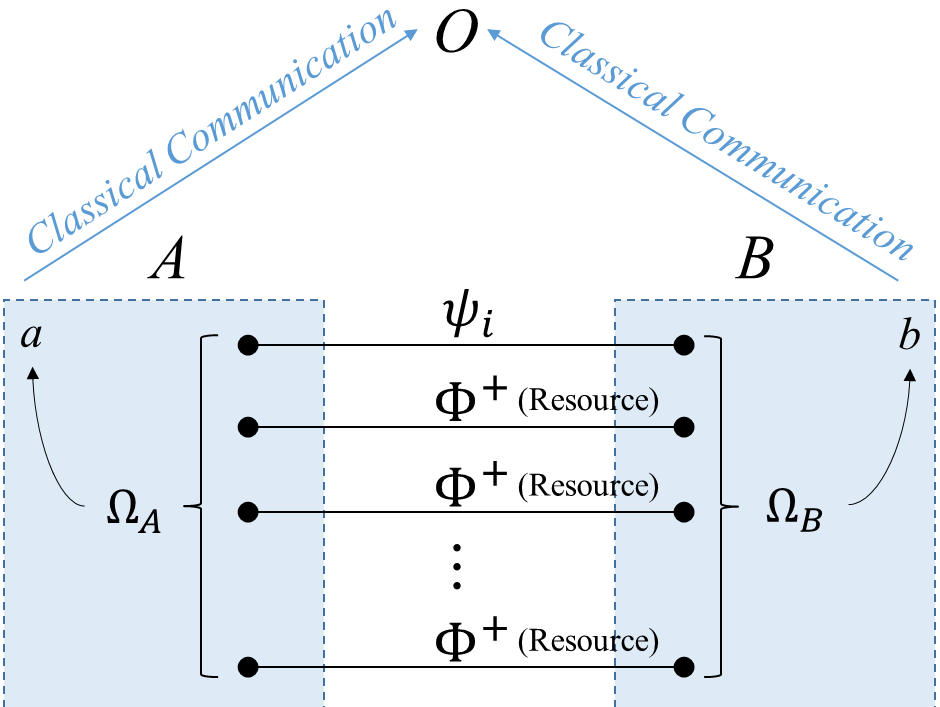}
    \caption{General scheme on how to implement bipartite localizable measurements with the aid of shared maximally entangled states (ebits) as a resource. To reconstruct the measurement result of a nonlocal measurement, an observer $O$ needs to put together the outcomes which are separately produced in the local laboratories $A$ and $B$.}
    \label{figloca}
\end{figure}

To begin with, by \textit{measurement} on a system prepared in a state $\rho$ we simply mean a quantum process that (i) produces a piece of classical information (outcome)  corresponding to an eigenstate $a_i$ of the operator associated to the measured observable $A$, and (ii) the probability to obtain $a_i$ is given by the Born rule: $ {\text{Prob}}(\Pi_i)= \text{tr} (\rho \Pi_i)$, where $\Pi_i$ is the projector onto the $i$-th eigenspace of $A$.  Note that a quantum measurement is completed at the moment when the classical outcome is produced, even if the readout for a single observer may require to collect (through classical communication) different pieces of classical information (see also Fig. \ref{figloca}). We also emphasize that the way the state is updated after the measurement is not part of the measurement itself.  
%%%%%%%%%%%%%%%%%%%%%%%%%%%%%%%%%%%
%%%%%%%%%%%%%%%%%%%%%%%%%%%%%%%%%%%

A measurement is then said to be \textit{ideal} if, when carried out again after an arbitrary small amount of time, it yields the same outcome $a_i$ with probability 1. This happens if the measurement process updates the state as (L\"uders rule): $\rho \rightarrow \rho'= \frac{\Pi_i \rho \Pi_i}{\text{tr} (\rho \Pi_i)}$, i.e., the post-measurement state lies within the eigen-subspace corresponding to the outcome. For non-degenerate operators, ideal measurements are projective. Here, for simplicity, we identify ideal and projective measurements. Note that in textbook quantum mechanics measurements are usually assumed to be ideal, but we will see that this can be problematic. 

It ought to be remarked that quantum measurements as defined by (i)-(ii) above do not define any \textit{measurement channel}. In fact, measurement channels also require to have post-measurement states, in which case one merely ignores (i.e., traces over) the measurement outcomes. \textit{Ideal measurement channels}, in particular,  correspond to ideal measurements where one ignores the outcome and only keeps the post-measurement states (with appropriate probability amplitudes that satisfy the Born rule). In this way, ideal measurement channels merely diagonalize the density matrix in the measurement basis but yield no outcome, i.e, they are not properly measurements.\footnote{In Ref. \cite{beckman2001causal}, the authors, by diagonalizing the density matrix to perform the ``measurement" are tacitly assuming an ideal state-update.}

Degenerate measurements can be realized by merely grouping outcomes. However, by doing so, an ideal non-degenerate measurement may turn into a non-ideal degenerate measurement.

A  measurement is said to be \textit{joint} if it is carried out on two (or more) subsystems, i.e., if the corresponding operator lives in the Hilbert space composed of the tensor product of the Hilbert spaces describing each of the subsystems.\footnote{Note that joint measurements formally also include product ones. Despite the fact that product measurements can trivially be carried out at a distance on their respective subspaces, for a single observer to know the result one still needs to classically communicate the local outcomes.} 
Non-relativistic quantum mechanics on its own does not impose any limitations on the possibility of measuring any observable, in our specific case, in the laboratory $M$ -- i.e., any Hermitian operator $\Omega$ living in $\mathcal{H}_M$ -- since the Hilbert space structure does not encompass the spatial  separation of the two qubits in $S$ and $R$.

In a real life experiment, to perform a joint measurement on both qubit systems one would bring them together and locally measure both. However,  in order to claim that the region $M$   is an extended region, as it is necessary for an impossible measurement scenario, we assume that the two qubits in $M$ cannot be brought together. A joint measurement in which the two subsystems cannot be physically brought together to perform it is said to be  \textit{nonlocal} (this is a physical constraint that is not encapsulated by the formalism of quantum mechanics). 

A nonlocal measurement is called \textit{localizable} if it can be carried out  by quantum operations performed on each of the parties without the possibility of signaling to each other (to reveal the result of the measurement one could need additional  classical communication, but the quantum operations that determine the results can be locally performed at two arbitrarily close time instants).\footnote{In relativity, this means that it exists a foliation such that the two events are arbitrarily close in time, possibly simultaneous.} We assume that to perform nonlocal measurement the parties can resort to some preshared resources, such as entanglement. A general scheme to implement a localizable measurement is depicted in Fig. \ref{figloca}: The measurement of a nonlocal observable with initial state $\psi_i$ is carried out indirectly by locally performing quantum operations $\Omega_{A}$ and $\Omega_{B}$ on the qubits located at $A$ and $B$, respectively. This produces classical outcomes that, if subsequently collected by an observer ($O$), unambiguously discriminate between the eigenstates of the nonlocal observable, i.e., reveals the result of the nonlocal measurement. 
We stress that all the discussions on nonlocal measurements and their localizability are insensible to the application of local unitaries on $A$ and $B$. Therefore in the remainder of this paper we will always tacitly assume that the result apply on equivalent classes of measurements up to local unitary transformations.

We have now all the ingredients to cast the main question about which measurements are physically possible in this framework, that is: 
\textit{What nonlocal joint measurements are localizable (i.e., can be implemented with local operations on spatially separated systems without violating no-signaling)?
}

%%%%%%%%%%%%%%%%%%%%%%%%%%%%%%%%%%%%%%%%

\section{An example of impossible measurements}
\label{example}
As already remarked in section \ref{imp}, the whole problematic step of the impossible measurement scenarios (both non-relativistically and in QFT) is the measurement carried out in the intermediate extended region $M$, because in any other step communication is allowed. Therefore, in the remainder of this paper, we will focus on the measurements at the middle region $M$ only, modeled, in the non-relativistic scenario,  by the  two separate qubits $A$ and $B$.

Let us start by showing that ideal nonlocal measurements lead in general to signaling. Following        \cite{groisman2001nonlocal, vaidman2003instantaneous}, consider the observable $T$ whose eigenstates are the orthogonal product states but twisted, i.e., they form the basis $\{\ket{00}, \ket{01}, \ket{1+}, \ket{1-}\}$, where $\ket{00}=\ket{0}_A\otimes \ket{0}_B$ etc., and $\ket{\pm}=1/\sqrt{2}(\ket{0}\pm\ket{1})$. Assume that the initial state of the two qubits $A$ and $B$ is $\psi_i=\ket{00}$. Alice can then decide whether to do nothing or to flip her qubit (i.e., to apply the ``kick" unitary $\sigma_x^{(A)})$, while Bob subsequently always measures in the computational basis.

If Alice does nothing to her qubit, Bob finds the outcome ``0" with probability $1$. But when she applies $\sigma_x^{(A)}$, the state gets updated to either of $\ket{1+}$ or $\ket{1-}$, therefore the local measurement of Bob yields either of the outcomes ``0" or ``1", each with probability  $1/2$. This is clearly a violation of Eq. \eqref{nsnorel}, leading therefore to signaling. We emphasize that what leads to signaling -- and therefore to the problem of the impossible measurement -- is the standard state-update rule (or the projection postulate) which is necessary for ideal measurements.

\section{Localizable quantum measurements}
\label{localiz}

\subsection{Twisted measurements}
\label{twisted}

\begin{figure}[ht]
    \centering    \includegraphics[width=.48\textwidth]{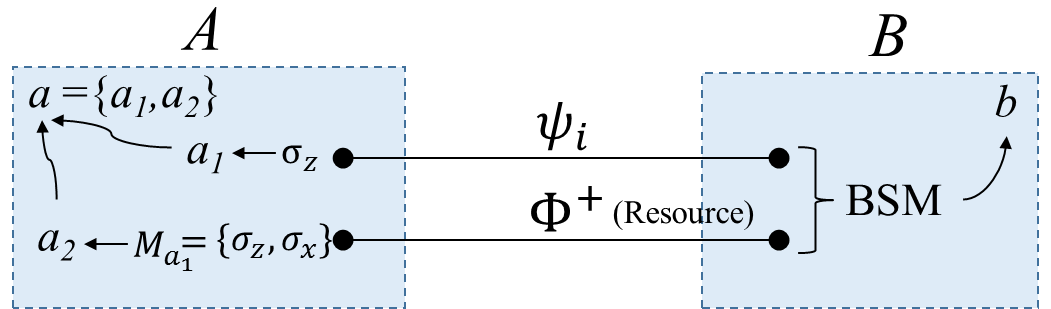}
    \caption{Scheme to implement a localizable measurement in a basis of  twisted product  states using an additional ebit as a preshared resource. $M_{a_1}$ is the measurement that Alice carries out on her second qubit, depending on the outcome $a_1$ of the measurement performed on her first qubit.}
    \label{fig3}
\end{figure}

We now show -- following Popescu, Groisman, Reznik, and Vaidman \cite{popescu1994causality, groisman2002measurements, vaidman2003instantaneous} -- that  there are ways to carry out a non-local measurement of $T$ (defined in section \ref{example}) without violating no-signaling, i.e., a measurement of $T$ is localizable. To do so, assume that Alice and Bob share the initial state $\psi_i$  and an extra ebit as a resource, say $\Phi^+=1/\sqrt{2}(\ket{00}+\ket{11})$ (see Fig. \ref{fig3}). The task is here to find a scheme that allows one to discriminate any of the eigenstates of the observable $T$ defined above. As previously pointed out, we are interested here only in nonlocal measurements, i.e., we require $A$ and $B$ to be carried out at a distance. However, both Alice and Bob are allowed to perform local joint measurements on the respective two qubits that they locally control.

The protocol goes as follows. Bob, performs a Bell State Measurement (BSM)  on his two local qubits, obtaining the outcome $b=0,1,2,3$. This updates the state of the two qubits on Alice's side as $\mathds{1}\otimes\sigma_b\psi_i$, where $\sigma_0=\mathds 1$, and $\sigma_j$, with $j=1, 2, 3$ are the standard Pauli matrices. Alice, on the other hand, independently and therefore possibly before or after or at a time arbitrarily close to Bob's measurement, carries out a measurement of her first qubit in the  Z-basis. If she finds the outcome $a_1=0$ she measures her second qubit in the Z-basis as well; otherwise, if she finds $a_1=1$, she measures the second qubit in the X-basis. The second measurement of Alice (in either basis), yields a bit $a_2$. Let us indicate by $a=\{a_1,a_2\}$ the 2-bit outcome of Alice's measurements.
 This ends the quantum part of the measurement, for all the outcomes are now encoded into pieces of classical information. To read out the result of this nonlocal measurement, Alice and Bob need to bring together the classical outcomes they respectively obtained, by for example sending them to an observer $O$ (see Fig. \ref{figloca}; henceforth the classical communication part will be omitted in the figures).\footnote{More in detail, while Alice’s measurement outcomes $a_1$ and $a_2$ allow to distinguish between the two groups of states $\{\ket{00},\ket{01}\}$ and  $\{\ket{1+},\ket{1-}\}$, this alone does not uniquely define the result of the nonlocal measurement. It is Bob’s measurement outcome $b$ that tells when the observer $O$ (see top of Fig. \ref{figloca}), who receives both Alice’s and Bob’s measurement results, has to flip bit $a_2$ in order to get the final result.} The merit of this scheme is that the whole quantum part of the protocol is carried out while keeping Alice and Bob spatially separated (it is nonlocal) but does not violate no-signaling (it is localizable).\footnote{An alternative technique to carry out a localizable measurement of this observable -- equivalent in terms of resources (i.e., it also uses an extra ebit) -- was previously proposed in \cite{groisman2002measurements}.} However, this measurement is not ideal, for the post-measurement state is not an eigenstate of the measured observable $T$ (Bob's qubit ends up in a totally mixed state).

It was shown in \cite{groisman2002measurements, groisman2003instantaneous} that this scheme can be extended to implement localizable measurements in the general product bases of two qubits of the form $T_\alpha=\{\ket{00}, \ket{01}, \ket{1   \alpha}, \ket{1\bar{\alpha}}\}$, where $\ket{\alpha}=\text{cos}(\frac{\alpha}{2})\ket 0 + \text{sin}(\frac{\alpha}{2})\ket 1$, and $\ket{\bar \alpha}=\text{sin}(\frac{\alpha}{2})\ket 0 - \text{cos}(\frac{\alpha}{2})\ket 1$, with $0<\alpha<\pi$. This measurement is asymptotically localizable, i.e., it requires an arbitrarily large number of ebits as resources to implement it. More precisely, for $n$ additional ebits, the probability of successfully carrying out a measurement of $T_\alpha$ is in general $1-1/2^n$. However, if $\alpha=k\pi/2^n$, only $n$ additional ebits are a sufficient resource to carry out a nonlocal measurement of $T_{\alpha}$. Hence, for any rational multiple of $\pi$, the number of resources needed to implement a localizable measurement is finite, and in particular can be very small, e.g., for $\alpha=\pi/4, \frac{3}{4}\pi$, $n=3$; for $\alpha=\pi/8, \frac{3}{8}\pi, \frac{5}{8}\pi$, $n=4$, and so on  \cite{groisman2002measurements}. 

%%%%%%%%%%%%%%%%%%%%%%%%%%%%%%%%%%%
%%%%%%%%%%%%%%%%%%%%%%%%%%%%%%%%%%%

Finally, based on numerical evidence, we formulate the following:
\begin{conj}
    The only localizable measurements on two qubits that can be performed using one ebit as a resource, and not less, are the twisted measurement, and the BSM.
\end{conj}

\subsection{Bell State Measurements}
\label{bsm}
\begin{figure}[ht]
    \centering
    \includegraphics[width=.48\textwidth]{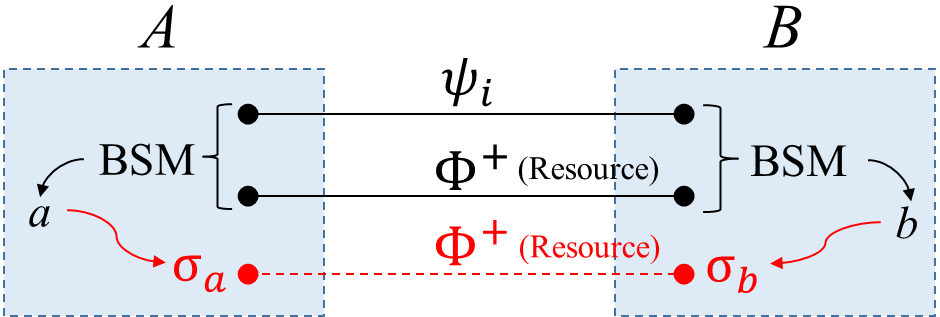}
    \caption{Scheme to implement a localizable BSM using an additional ebit as a resource. If a further ebit (represented as dashed, in red) is preshared, the measurement can be ideal. BSM is the only (non-product) measurement of two qubits that is localizable and ideal.}
    \label{fig4}
\end{figure}

The most well known joint (entangled) measurement is surely the BSM, i.e., the observable whose eigenvectors are the four Bell states $\{\Psi^\pm, \Phi^\pm \}$. BSMs are localizable with the aid of only one extra ebit (Fig. \ref{fig4}). The pair of qubits on which they want to carry the nonlocal BSM is prepared in an initial state $\psi_i$. The protocol goes as follows. Both Alice and Bob perform BSM on their respective local qubits, each of them obtaining an outcome, $a=0,1,2,3$ and $b=0,1,2,3$, respectively. These correspond to the four possible outcomes of the measurements, associated, respectively, to the eigenstates of the BSMs $\Phi^+$, $\Psi^+$, $\Psi^-$, and $\Phi^-$. 

It is well-known that for BSMs performed on a pairs of qubits distributed along each of the edges of a polygon, here a rectangle (the black one in Fig. \ref{fig4}), the amount of ``$\Phi$"'s (alternatively ``$\Psi$"'s) and the amount ``+"'s (alternatively ``-"'s) should be even. Therefore, if the resource ebit is prepared in a fixed Bell state, say again $\Phi^+$, the outcomes $a$ and $b$ fully determine the result of the BSM. For example, if Alice finds $a=2$, corresponding to the eigenstate $\Psi^-$, whereas Bob finds $b=3$, corresponding to $\Phi^-$, then the result corresponds to the eigenvalue associated to $\Psi^+$, with probability $\text{Prob}(a=2,b=3)=|\braket{\psi_i}{\Psi^+}|^2$. An observer collecting the classical pieces of information $a$ and $b$ can therefore figure out the result of the nonlocal measurement. 
%%%%%%%%%%%%%%%%%%%%%%%%%%%%%%%%%%%
%%%%%%%%%%%%%%%%%%%%%%%%%%%%%%%%%%%

This scheme shows that BSM is therefore localizable with the use of an additional ebit, but it is not ideal, i.e., the post-measurement state is not prepared in an eigenstate of the measurement operator.\footnote{Note that also partial BSM can be performed with this scheme, although not ideally.} However, one can consider to use another ebit as a resource (dashed, in red in Fig. \ref{fig4}). In that case, depending on the output of their local BSMs, Alice and Bob apply, to their respective part of the second ebit, the local operators $\sigma_a$ and $\sigma_b$, respectively. In this way, they end up with a shared post-measurement state that is indeed the measured eigenstate, rendering the measurement ideal as well as localizable. This means that if they repeat the measurement procedure immediately afterwards (by using another extra ebit), although the single local outcomes $a$ and $b$ may differ, they combine to yield again the same result.

In fact, in Ref. \cite{popescu1994causality}, it was proven that BSM is the only measurement of two qubits that is at the same time localizable and ideal (besides the trivial case of the product observables). For the degenerate case, the same result about localizabily holds, although the measurement cannot be ideal. The intuitive reason is that BSM is special because it completely erases the information about the pre-measurement states, which is a necessary condition to implement an ideal localizable measurement \cite{popescu1994causality} (see also section \ref{general}). 

Let us emphasize that the BSM is an exceptional case, and should thus not be taken as a general example of joint measurement.

%\textcolor{blue}{GHZ? I would only mention them here (or even only in the next general section, lemma 2) and provide any detail in appendix.}

\subsection{Elegant Joint Measurements}
\label{ejm}

The Elegant Joint Measurement
(EJM) is an observable whose eigenvectors form a basis of partially entangled states (all with the same degree of entanglement) with particular symmetry properties
\cite{gisin2019entanglement} (see also \cite{del2023iso} for a classification of joint entangled measurements of two qubits). Indeed, their reduced states point at the vertices of the regular tetrahedron inscribed in the Bloch sphere. The states assume the functional form: $\Phi_j=\frac{\sqrt{3}+1}{2\sqrt{2}}\ket{\Vec{m}_j,-\Vec{m}_j} +\frac{\sqrt{3}-1}{2\sqrt{2}}\ket{-\Vec{m}_j,\Vec{m}_j}$, where $\ket{\Vec{m}_j,-\Vec{m}_j}$ means $\ket{\Vec{m}_j}\otimes\ket{-\Vec{m}_j}$ and $\Vec{m}_j$ are the vectors pointing at the vertices of the tetrahedron \cite{gisin2019entanglement}. %EJM of two qubits  play a special role in network nonlocality \cite{gisin2019entanglement, tavakoli2021bilocal}.
%%%%%%%%%%%%%%%%%%%%%%%%%%%%%%%%%%%
 \begin{figure}[ht]
    \centering
    \includegraphics[width=.48\textwidth]{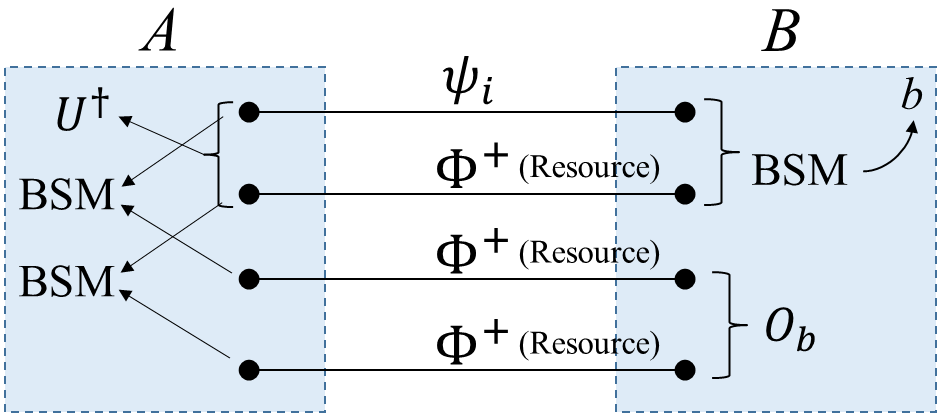}
    \caption{Scheme to implement a localizable EJM using three  additional ebits as a resource. On Alice's side, first the unitary $U^\dagger$ is applied to the first two qubits, then BSMs are performed between the qubits 1-3 and 2-4.}
    \label{fig5}
\end{figure}
%%%%%%%%%%%%%%%%%%%%%%%%%%%%%%%%%%%

We now show that nonlocal EJMs are localizable using three ebits as a resource. With reference to Fig. \ref{fig5}, assume that $\psi_i$ is one of the eigenstates of the EJM observable. All the three additional ebits are prepared in a $\Phi^+$ Bell state. As in Fig. \ref{fig5}, Alice first applies the operator $U^\dagger$ on the first pair of qubits, where $U$ is the unitary operator whose columns are the four EJM states. Then she performs two BSMs on the two pairs of qubits 1-3 and 2-4, obtaining the outcomes $a_1$ and $a_2$, respectively. Bob, instead, performs a BSM on two of his qubits and, depending on the the outcome $b$ he obtains, he measures the other two qubits applying $O_b$ defined as following. If $b=0$, this corresponds to a measurement in the Z-Z product basis $\{\ket{+z,+z}, \ket{+z,-z}, \ket{-z,+z} \ket{-z, -z}\}$; if $b=1$, in the Y-Z Bell basis $\{\ket{+y,+z}\pm \ket{-y, -z},\ket{+y,-z}   \pm \ket{-y, +z}\}$; if $b=2$, in the Z-Y Bell basis $\{\ket{+z,+y}\pm \ket{-z, -y},\ket{+z,-y}    \pm \ket{-z, +y}\}$; finally, if $b=3$, he measures in the Y-Y product basis $\{\ket{+y,+y}, \ket{+y,-y}, \ket{-y,+y} \ket{-y, -y}\}$. Here $\ket{\pm y}$ and $\ket{\pm z}$ are the two possible eigenstates of a qubit measured in the $Y$ and in the $Z$ direction, respectively.  The result of the nonlocal measurement can be retrieved by an observer (after the classical communication) via the expression
\begin{eqnarray}
\label{ejmstate}
\psi_{B_2}=\sigma_{a_1}\otimes \sigma_{a_2} U^{\dagger}\mathds{1}\otimes\sigma_{b}\psi_i, %\nonumber
\end{eqnarray}
where $\psi_{B_2}$ is the state of the two qubits of Bob to which he applies $O_b$ (i.e., the two bottom right in Fig. \ref{fig5}). EJMs are therefore localizable using three additional ebits (but cannot be ideal).

Based on numerical evidence, we put forward the following:
\begin{conj}
    EJM are essentially the only localizable measurement of two qubits with entangled eigenstates that can be measured with three ebits (and no less) as resources.\footnote{In fact, besides the EJM, we find three further bases that can be related to EJMs. We are investigating this in further detail and a manuscript is in preparation.}
\end{conj}

%\textcolor{blue}{How to include your "almost the only measurments"?}

\subsection{Review of general results}
\label{general}
In this subsection we summarize some general results presented above and from the literature.
\begin{theorem}
Every localizable measurement must completely erase all local information  about the pre-measurement state in order to be ideal.
\end{theorem}
\noindent The proof is provided in Ref. \cite{popescu1994causality}.
%%%%%%%%%%%%%%%%%%%%%%%%%%%%%%%%
\begin{lemma}
        The only localizable and ideal nonlocal measurement of two qubits is the BSM (and product measurements).
\end{lemma}
\noindent The proof is provided in Ref. \cite{popescu1994causality}. In section \ref{bsm}, we explicitly showed that to implement a localizable BSM one requires an extra ebit as a resource, but to implement it ideally one needs two ebits. Moreover, for N-partite systems on can prove:
\begin{lemma}
        GHZ measurements on $N$ qubits are localizable and ideal.
        \end{lemma}
\noindent A GHZ measurement is defined as the observable whose eigenstates form the (n-partite) GHZ basis (see, e.g, \cite{de2020protocols}).
This lemma follows as a trivial multipartite extension of the construction for BSMs in \ref{bsm}.
%%%%%%%%%%%%%%%%%%%
%\textcolor{blue}{Do we have any idea about Werner states? Is GHZ unique like Bell?}

The main result concerning localizability is the following:
%%%%%%%%%%%%%%%%%%%%%%
\begin{theorem}
        All nonlocal measurements (of any dimension and any number of parties) are localizable.
\end{theorem}
\noindent The constructive proof is provided in Ref. \cite{vaidman2003instantaneous}. As already noted, in general these measurements cannot be ideal and their implementation  requires an infinite amount of resources. However, one can show:
\begin{lemma}
    There are localizable measurements that require only finite entangled resources.
\end{lemma}
\noindent Concrete examples are provided in sections \ref{twisted}, \ref{bsm}, and \ref{ejm}.

In conclusion, given enough resources, “impossible” measurements are always localizable (Theorem 2), hence physically possible, but cannot in general be ideal (Theorem 1).
%%%%%%%%%%%%%%%%%%%%%%%%%%%%%%%%%%%%%
%%%%%%%%%%%%%%%%%%%%%%%%%%%%%%%%%%%%%
\section{Discussion and outlook}
%%%%%%%%%%%%%%%%%%%%%%%%%%%%%%%%%%%%%
%%%%%%%%%%%%%%%%%%%%%%%%%%%%%%%%%%%%%
We have shown (i) that Sorkin's impossible measurements, initially raised as a problem of QFT, have a non-relativistic limit as quantum nonlocal measurements that lead to signaling between separated parties. They should therefore be addressed as a general foundational  problem of quantum physics, independently of QFT. We then showed (ii) that the problem of impossible measurements is solved (in non-relativistic quantum mechanics) by considering localizable measurements of nonlocal variables. That is, those measurements that can be carried out at a distance without violating no-signaling, and which we deem, therefore, physically possible. Moreover, (iii) we have recalled that almost all localizable measurements are non-ideal, i.e., are not immediately reproducible. We emphasize that from this point of view the well-known BSM is exceptional and, thus, should not be used to build our intuition about joint measurements. Finally, (iv) we reviewed and developed further the theory of localizable measurements, bridging the gap between results put forward by different communities that had, seemingly, not been connected so far. 

%We have discussed how this provides a solution to the non-relativistic limit of .
Sorkin's impossible measurements in QFT have recently received a great deal of attention, mostly thanks to a proposed solution that, at last, puts forward an explicit measurement scheme for QFT \cite{fewster2020quantum, fewster2020generally, bostelmann2021impossible, fewster2023measurement}. %That approach, however,  excludes by construction the possibility of having genuine nonlocal measurements, for it locally couples the fields to a probe (a quantum field itself) that then subluminally mediates the interactions within the region \textit{M} (see also \cite{van2021relativistic}).\footnote{This is equivalent to a semilocalizable measurement in the parlance of \cite{beckman2001causal, eggeling2002semicausal}, i.e., one that can be carried out by sending one-way communication between the parties. Similar schemes were also developed in \cite{groisman2001nonlocal}.} While this is certainly a viable solution, it has the risk of being too restrictive in ruling out \textit{a priori} the physicality of nonlocal measurements. 
%In our modeling of the non-relativistic limit, this would correspond, to introduce another system that couples to either of the two systems $A$ and $B$, which physically travels from one to the other. 
That approach, however,  has so far not addressed explicitly the possibility of measuring nonlocal variables, and only considered the fields to be measured by coupling them to a probe (a quantum field itself) that then subluminally mediates the interactions within the extended region \textit{M} (see also \cite{van2021relativistic}).\footnote{This is equivalent to a semilocalizable measurement in the parlance of \cite{beckman2001causal, eggeling2002semicausal}, i.e., one that can be carried out by sending one-way communication between the parties. Similar schemes were also developed in \cite{groisman2001nonlocal}.} We believe that by introducing entangled probes all the results here presented could be eventually reproduced in the language of algebraic QFT, but this remains to be done. % The non-relativistic construction here presented could hopefully inspire QFT to achieve a complete theory of measurement . 
%In our modeling of the non-relativistic limit, this would correspond, to introduce another system that couples to either of the two systems $A$ and $B$, which physically travels from one to the other. 
These considerations can thus perhaps bring novel insights in helping guide the development of a complete theory of measurements in QTF, i.e. one that yields outcomes and prescribes how to update the state afterwards.

Moreover, even in standard quantum mechanics, joint measurements remain poorly understood, an issue that has recently been listed among the most relevant open problems in the foundations of quantum physics   \cite{cavalcanti2023fresh}. Only lately, some attempts to systematize joint measurements have been put forward \cite{chitambar2014everything, del2023iso}, but much has still to be done. The results surveyed in this paper show that with enough resources, every nonlocal measurement is localizable, although it is almost never ideal. We further showed several cases in which the amount of resources is finite and relatively small, making this proposal physically compelling. This suggests that localizability could help characterize joint measurements by the necessary amount of resources (i.e. entanglement) that they require. Such an approach can single out some measurements (or classes thereof) with special proprieties.\footnote{Some work aimed to characterize the amount of resources need for localizability was already provided in Refs. \cite{clark2010entanglement, groisman2015optimal}, but only in the asymptotic limit} For instance, BSM is the only possible measurement (up to local unitaries) to be localizable and ideal. Perhaps other measurements, like the EJM, can be singled out in a similar way (cf. Conjecture 2). We are addressing this problem more systematically and a manuscript addressing the classification of joint measurements by their property of localizability is currently in preparation.

A further interesting question would be to analyze Sorkin's impossible measurements in different theoretical frameworks. This could help understand the minimal requirements for a theory to exhibit this kind of problems, namely, what is fundamentally the source of impossible measurements? This paper shows that this is not a problem of relativistic quantum physics per se. One might then think that the source may be nonlocality, but the most nonlocal of the no-signaling (fictional) theories, so-called ``boxworld", exclusively admits separable joint measurements \cite{eftaxias2022joint} and therefore does not have a Sorkin's problem. 
%The results here reviewed point at the fact that the standard state-update rule plays a major role. This may lead to attempts to find different state-updates after measurements \cite{fiorentino2023quantum}. 
On the other hand, a recent work claims that impossible measurements can be found in a classical limit (i.e., non-quantum) of QFT \cite{much2023superluminal}, also seemingly ruling out that this is a genuine quantum effect.

Finally, in the axiomatic approach to quantum theory based on property-lattices \cite{jauch1969structure}, Piron’s theorem states that, assuming certain natural axioms, like the covering law (the projection of a pure state is necessarily a pure state), the lattice of quantum properties is isomorphic to the closed subspaces of a linear space, possibly with superselection rules \cite{piron1964axiomatique}. One of us, NG, proved that the lattice of non-local properties is atomic (it has pure states) \cite{gisin1986property}. Here we proved that it is also ortho-modular (each property has an orthogonal one). But since we showed that (almost all) nonlocal measurements cannot be ideal, the covering law is a doubtful axiom. As it is the latter that ensures linearity in quantum physics, could this be a hint that one should relax linearity in QFT?

\acknowledgements
We thank Sandu Popescu, Daniel Collins, 
Christopher Fewster, Tein van der Lugt, Kuntal Sengupta, Mirjam Weilenmann, and Pavel Sekatski for useful discussions and for pointing out relevant references. 
This research was supported by the FWF (Austrian Science Fund) through an Erwin Schr\"odinger Fellowship (Project J 4699), and the Swiss National Science Foundation via the NCCR-SwissMap.

\bibliography{biblio}

\end{document}